\documentclass[twocolumn,prl,showpacs,preprintnumbers,superscriptaddress,amsmath,amssymb]{revtex4} 
\usepackage{graphicx}
\usepackage{color}
\usepackage{bm}
\newcommand{\BE}{\begin{equation}}
\newcommand{\EE}{\end{equation}}
\def\BEA{\begin{eqnarray}}
\def\EEA{\end{eqnarray}}

\def\on{{O_{n}}}

\def\cpla{{\lambda_{a}}}
\def\cplb{{\lambda_{b}}}
\def\cplc{{\lambda_{c}}}

\def\cpln{{\lambda_{n}}}
\def\cple{{\lambda_{\epsilon}}}

\def\const{{\Omega}}
\begin{document}
\title{Spin liquid versus dimer phases in an anisotropic $J_1$-$J_2$
 frustrated square  antiferromagnet}
\author{Alexandros Metavitsiadis}
\affiliation{Department of Physics, University of Kaiserslautern,
D-67663 Kaiserslautern, Germany}
\author{Daniel Sellmann}
\affiliation{Department of Physics, University of Kaiserslautern,
D-67663 Kaiserslautern, Germany}
\author{Sebastian Eggert}
\affiliation{Department of Physics, University of Kaiserslautern,
D-67663 Kaiserslautern, Germany}

\date{\today}
\vskip\baselineskip
\begin{abstract}
The spin-1/2 $J_1$-$J_2$ antiferromagnet is a prototypical model 
for frustrated magnetism and one possible candidate for a realization of a spin liquid
phase.  The generalization of this system on the anisotropic square lattice is
given by the $J_1$-$J_2$-$J_1'$-$J_2'$ Heisenberg model, which can be treated by
a renormalization group (RG) analysis of coupled frustrated chains.
The $J_1$-$J_2$-$J_1'$-$J_2'$-model
shows an interesting interplay of N\'eel order, dimerization, and spin liquid behavior.
The analytical findings supported by numerical results indicate that for the isotropic 
model the phase at intermediate coupling strength $0.4 J_1 \alt J_2 \alt 0.6 J_1$ is not 
a spin liquid, but has instead a spontaneously staggered dimer order.  
\end{abstract}

\pacs{75.10.Jm, 75.10.Kt, 75.30.Kz}
%71.10.Hf Non-Fermi-liquid ground states, phase transitions in model systems
% 67.80.kb Supersolid phases on lattices
%74.25.Nf Response to electromagnetic fields (nuclear magnetic resonance
%75.10.-b General theory and models of magnetic ordering 
%75.10.Dg Crystal-field theory and spin Hamiltonians 
%75.10.Hk Classical spin models
%75.10.Jm Quantized spin models, including quantum spin frustration
%75.10.Kt Quantum spin liquids, valence bond phases and related phenomena
%75.10.Lp Band and itinerant models
%75.10.Nr Spin-glass and other random models
%75.10.Pq Spin chain models
%75.20.Hr Local moment in compounds and alloys; Kondo effect
%75.25.+z Spin arrangements in magnetically ordered materials
%75.30.Ds Spin waves (for spin-wave resonance, see 76.50.+g)
%75.30.Hx Magnetic impurity interactions
%75.30.Mb Valence fluctuation, Kondo lattice, and heavy-fermion phenomena
%75.30.Hx Magnetic impurity interactions
%75.30.Kz Magnetic phase boundaries
%75.30.Sg Magnetocaloric effect, magnetic cooling
%75.40.-s Critical-point effects, specific heats, short-range order
%75.40.Cx Static properties (order parameter, static susceptibility,..
%75.40.Mg Numerical simulation studies
%75.50.Ee Antiferromagnetics
%76.60.-k Nuclear magnetic resonance and relaxation
%76.60.Cq Chemical and Knight shifts

\maketitle
{%
In the search of exotic quantum states and quantum phase transitions,
frustrated antiferromagnetic systems have increasingly become the center of
attention \cite{spin-liquid, balents-nature}.  
The competing interactions between spins
potentially lead to a large entropy even at low temperatures, which together with
quantum fluctuations may give rise to quantum phases with unconventional or topological 
order parameters \cite{NP.8.902,PhysRevLett.96.110404, PhysRevLett.96.110405}. 
In particular, the so-called spin liquid state
without long range order of a conventional (local) order parameter
has been much discussed in the literature ever since Anderson related this phase to
high-temperature superconductivity \cite{anderson}.
A solid proof for a system which shows a spin liquid ground state has long been 
elusive, due to inherent numerical and analytical problems in frustrated systems.
Nonetheless, some good evidence for possible spin liquid states has recently 
been presented 
for the Hubbard model on the honeycomb \cite{honeycomb-nature} and the 
anisotropic triangular lattice \cite{tocchio}, as well as for
for the Heisenberg
model on a Kagome lattice \cite{spin-liquid,jiang,white-science,becca2}, 
and on the 
$J_1$-$J_2$ frustrated square lattice \cite{balentsj1j2,becca1,gong}. 
\par 
In particular, the $J_1$-$J_2$ Heisenberg model has been treated with a vast array of theoretical 
methods \cite{balentsj1j2,becca1,gong,schulz,PhysRevB.42.8206,PhysRevLett.93.127202,PhysRevB.81.144410,PRB_79_024409,PRB_44_12050,PRB_85_094407,PRB_74_144422,sirker2006,PRL_91_197202,PhysRevB.78.214415,1309.6490,PRL_91_067201,PRB_86_045115,PRL_111_037202,PhysRevB.63.104420,PhysRevB.66.054401,PRB_41_9323,PRB_54_9007,PRL_78_2216,PRB_60_7278,PhysRevLett.91.017201,PRL_84_3173,PRL_87_097201,EPJB_73_117,PhysRevB.58.6394,1112_3331,PhysRevLett.97.157201,Senthil}, which is also the model originally considered by  Anderson \cite{anderson}.  
Early numerical works have shown 
an intermediate phase between ordinary N\'eel order for $J_2 \alt 0.4J_1$ and 
collinear N\'eel order for $J_2 \agt 0.6J_1$ \cite{schulz}, but the underlying 
correlations in this phase remain hotly debated.  
Most works have predicted a plaquette or columnar dimer
order as the most likely scenario \cite{PhysRevB.42.8206,PhysRevLett.93.127202,PhysRevB.81.144410,PRB_79_024409,PRB_44_12050,PRB_85_094407,PRB_74_144422,sirker2006,PRL_91_197202,PhysRevB.78.214415,1309.6490}, but more recent numerical works have again proposed a spin liquid \cite{balentsj1j2,becca1,gong}. Unfortunately, 
the issue may 
never be conclusively solved using numerical methods, since 
%the $J_1$-$J_2$-model cannot be treated
%with Quantum Monte Carlo methods and at the same time 
convergence with finite size and/or
temperature of data from density matrix renormalization group (DMRG) or tensor matrix methods
is very slow.  
In particular, it was shown recently for the related two-dimensional (2D)
$J$-$Q$ model that the finite size scaling on moderate lengths 
would lead to the prediction of a spin liquid, even though 
the system orders in the thermodynamic limit \cite{sandvik2012}.
In general, the  slow convergence
is related to the observation that spin liquid states are often very close in energy to
competing states with dimer order \cite{white-science,becca2}.
\begin{figure}[t!]
\includegraphics[width=0.85\columnwidth]{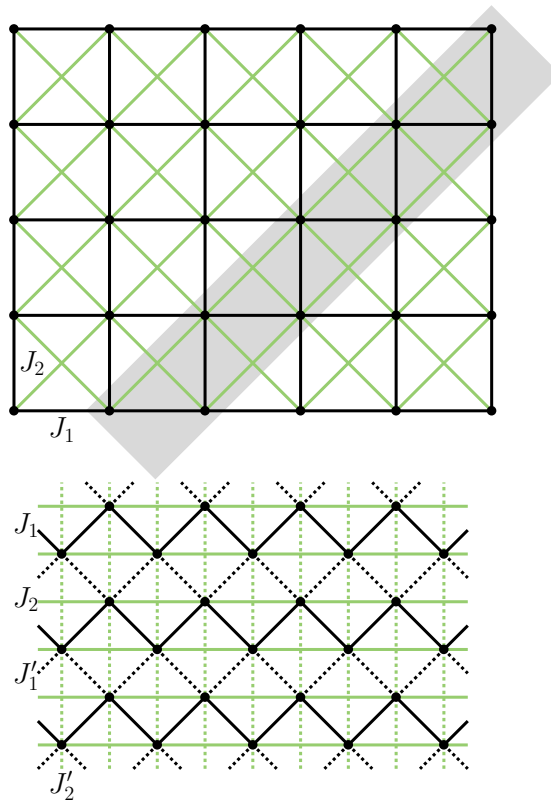}
\caption{The $J_1$-$J_2$ square antiferromagnet (top) and the generalized 
anisotropic $J_1$-$J_2$-$J_1'$-$J_2'$ antiferromagnet obtained from a diagonal 
slice of the square antiferromagnet (bottom).}\label{model}
\end{figure}

%Frustrated spin systems are difficult to handle numerically since quantum Monte
%Carlo simulations suffer from the infamous minus sign problem.  Numerical methods
%based on the density matrix renormalization group (DMRG) open another 
%possibility \cite{white-science,balentsj1j2}, but in two dimensions
%the convergence to the true ground state often turns out to be extremely slow with
%system size and/or temperature.  In particular, it was shown for the so-called $J$-$Q$ model
%that the finite size scaling is very slow \cite{sandvik2012}, which was taken as an 
%indication that 
%DMRG results for the frustrated $J_1$-$J_2$-model \cite{balentsj1j2} are not 
%yet conclusive evidence for a spin liquid ground state \cite{sandvik2012}.
%In general, the  slow convergence 
%is related to the observation that spin liquid states are often very close in energy to 
%competing states with dimer order \cite{white-science,becca2}. 
%
\par 
In light of the disappointing numerical situation, analytical methods become very
important, but the problem is difficult since 
series expansions \cite{PhysRevB.42.8206, PhysRevLett.91.017201}, chain mean-field theories 
\cite{PhysRevB.72.094416}, spin wave 
theories \cite{sirker2006} or
coupled cluster methods \cite{PhysRevB.78.214415, PhysRevB.58.6394, PhysRevLett.97.157201}
 have to incorporate frustrating interactions, which tend to 
cancel each other.  Therefore, the methods often become unstable or diverge near the most 
interesting phase transitions.  
%In particular, for the $J_1$-$J_2$-model different
%methods may give contradictory results about the nature of the 
%phase transitions \cite{sirker2006, PhysRevB.78.214415, Senthil05032004, PhysRevB.81.144410,
% PhysRevB.63.104420, PhysRevB.66.054401}.  
Another very promising analytical approach 
is to couple one-dimensional (1D) systems \cite{PhysRevLett.93.127202, giamarchi,
 PhysRevLett.98.077205,  PhysRevB.67.024422, science.271.618}. 
However, 
even in this case the frustrating inter-chain interactions tend to cancel each other 
so that interesting new phases only come up in higher order 
treatments at intermediate inter-chain coupling strength.
This is not necessarily a fundamental limitation, 
however, since the leading frustration effects can be taken into account 
within each chain if a different geometry is chosen as we will show here.
\par 
We now propose to study a more general $J_1$-$J_2$-$J_1'$-$J_2'$-model, which 
takes into account the effects of frustration {\it within} 
each chain before they are coupled, but at the same time allows to make 
exact analytic predictions in parts of the phase diagram.
The model is given by the standard %SU(2) invariant
Heisenberg model on the 2D lattice 
\begin{eqnarray}\label{hamiltonian}
H & = &  \sum_{\langle i,j \rangle}J_{ij}  \mathbf{S}_{i} \cdot \mathbf{S}_{j}
\end{eqnarray}
with $J_{ij}>0$ acquiring values according to the geometry of Fig.~\ref{model}.
Notice that the lattice vectors of the square 
lattice are tilted by $45^{\rm o}$ relative to the 
directions of the zig-zag chains.
%nearest- and next-nearest-neighbor interaction as depicted in 
One special property of this model is that it can be treated by a 
renormalization group (RG) treatment of quasi 1D chains which becomes exact
in the limit of small interchain coupling $J_1'$, $J_2'$ while  
the in-chain frustration $J_2$ is now {\it finite} and
can be fully taken into account without the need to go to higher order RG equations.
In fact, it turns out that $J_2$ becomes the determining factor 
for the behavior of the phase diagram.  We predict 
the intermediate phase to posses a {\it staggered} dimer order, which is in contrast
to previous considered states of spin liquids \cite{balentsj1j2, becca1}, 
columnar dimer order, and plaquette dimer order \cite{sirker2006,PhysRevB.78.214415, 
PhysRevLett.93.127202}.  
}

\par 
Our starting point is the 1D model 
without inter-chain couplings $J_1'=J_2' = 0$, which  is well understood in terms
of a 1+1 dimensional field theory Hamiltonian based on Abelian or non-Abelian 
bosonization \cite{Ian1986409}.
In the continuum limit the spin operators are
expressed in terms of field theory operators
\BE \label{spin operator}
\mathbf{S}(x)\approx
\mathbf{J}(x)+(-1)^{x}~\const~\mathbf{n}(x)~,
\EE
where $\const$ a non-universal constant of order one,
  $\mathbf{J}=\mathbf{J}_{L}+\mathbf{J}_{R}$ is 
the sum of the chiral $\mathrm{SU(2)}$ currents of the Wess-Zumino-Witten (WZW) model and $\mathbf{n}$ 
is related to the fundamental field $g$ of the WZW model via $\mathbf{n}=\text{tr} {\bm \sigma} g$.    The lattice constant is set to unity here.
The conformally invariant fixed point Hamiltonian $H_0$ of the frustrated chain is perturbed by   
the backscattering   marginal  operator \cite{JPhysA.22.511}
\begin{eqnarray}
H & = &  H_0+2 \pi v \cpla\int dx~ \mathbf{J}_L\cdot \mathbf{J}_R  , \ \ \ \ \text{with}\\
 \label{ham-single-chain}   
H_0 & = &  \frac{2\pi v}{3} \int dx \big[ :\!\mathbf{J}_L\cdot \mathbf{J}_L\!\!:+:\!\mathbf{J}_R\cdot \mathbf{
J}_R\!\!:\big]~, \nonumber 
\end{eqnarray}
where the dots denote normal ordering. The velocity is approximately $v=\tfrac{\pi J_1}
{2}-1.65J_2$ if a linear dependence on $J_2$ is assumed \cite{eggert92} while  
to leading order \cite{PhysRevB.54.R9612} $\cpla \approx 1.723(J_{2}-J_{c})$ where
$J_{c} \approx 0.241167J_1$ \cite{0305-4470-27-17-012,PhysRevB.54.R9612,PhysRevB.54.9862,PhysRevB.25.4925}.
\par 
The leading instabilities of the chains are a finite 
staggered field $(-1)^{x}\mathbf{S}(x)\propto\mathbf{n}$ with 
effective scaling dimension $d_n=1/2+\lambda_a/4$ and a dimerization 
$(-1)^{x}\mathbf{S}(x)\cdot \mathbf{S}(x+1) \propto  \epsilon$,
where $\epsilon = \text{tr} g $ has effective scaling dimension $d_\epsilon=1/2-3\lambda_a/4$
\cite{JPhysA.22.511}. The marginal coupling $\lambda_a$ itself renormalizes and becomes 
relevant for $\lambda_a>0$, which leads to a dimerized phase for $J_{c}< J_2  < \infty$
\cite{0305-4470-27-17-012,PhysRevB.54.R9612,PhysRevB.54.9862,PhysRevB.25.4925,PhysRevB.55.299}. 
Therefore, the in-chain frustration $J_2$ 
determines if the leading instability is dimerization for $J_2>J_c$ or N\'eel order
for $J_2<J_c$.  For $J_2=J_c$ the system is a spin liquid in form of a Luttinger liquid.

\par 
In the presence of inter-chain couplings $J'_{1}$ and $J'_{2}$ the field theories 
of neighboring chains become coupled and  give rise to a richer operator content 
\cite{PhysRevB.77.205121,PhysRevB.73.214427,PhysRevB.72.104435}.  
Any chain in the system is coupled to two neighboring chains, so it is 
natural to consider three copies of the field theory that are weakly coupled.
%according to the microscopic coupling in the $J_1$-$J_2$-$J_1'$-$J_2'$-model.
For self-consistency all three copies must renormalize in the same way, which implies
the following symmetric formulation of the perturbing Hamiltonian
in terms of three non-Abelian field theories $\nu=1,2,3$
\BEA\label{hpert}
&\delta \mathcal{H}&= 2 \pi v \sum_{\nu=1}^{3} \Big[\cpla 
\mathbf{J}_{\nu, L}\cdot \mathbf{J}_{\nu, R}
+  \cpln \mathbf{n}_\nu\cdot \mathbf{n}_{\nu+1} 
+\cple \epsilon_\nu \epsilon_{\nu+1} \nonumber \\
&+& \cplb(\mathbf{J}_{\nu, L}\cdot \mathbf{J}_{\nu+1, R} +\mathrm{e.c.}) +
\cplc (\mathbf{J}_\nu\cdot \mathbf{n}_{\nu+1}-\mathrm{e.c.})\Big],
\EEA
with $\mathrm{e.c.}$ denoting  the operator after exchanging the chain index,  
and the bare couplings are
\BE\label{barecouplings}
\cpln =\const^2 \frac{(J'_2-J'_1)}{2\pi v},~ \cplb =\frac{J'_1+J'_2}{2\pi v},~ 
\cplc=\const \frac{J'_1}{2\pi v},~ \cple=0.
\EE
\par 
Letting the ultra-violet cutoff evolve according to $\Lambda(l)=\Lambda_0 e^{-l}$ 
and using the operator product expansion \cite{book-senechal-2004,cardy},
we get the following coupled differential equations for the coupling constants
(see appendix)
\begin{subequations}\label{rg-flow-a}
\BEA
\dot{\cpla} \hskip-0.2cm &=& \hskip-0.2cm \cpla^2 + \cple^2 -
\cpln^2~, \label{rg-cpla} \\
\dot{\cplb} \hskip-0.2cm &=& \hskip-0.2cm  \cplb^2 -\cple\cpln
+ \cpln^2~,
\label{rg-cplb} \\
\dot{\cple} \hskip-0.2cm &=& \hskip-0.2cm \cple +\frac{3}{2}\cpla\cple
-\frac{3}{2}\cplb\cpln -\frac{3}{2}\cplc^2 ~, \label{rg-cple}\\
\dot{\cpln} \hskip-0.2cm &=& \hskip-0.2cm \cpln -\frac{1}{2}\cpla\cpln
-\frac{1}{2}\cplb\cple +\cplb\cpln -\cplc^2 ~, \label{rg-cpln}\\
\dot{\cplc} \hskip-0.2cm &=& \hskip-0.2cm \frac{1}{2}\cplc -\frac{1}{4}\cpla\cplc
+\cplb\cplc + \frac{1}{2}\cplc\cple -\cplc\cpln \label{rg-cplc}~ ,
\EEA
\end{subequations}
where $\dot{\lambda}=d\lambda/dl$.
% denotes differentiation with respect to $l$.
 
The system remains approximately scale invariant as long as $\Lambda(l)$ is
the largest energy scale in the system.
However, typically one of the coupling constants $\lambda$ becomes 
of order unity at one point so that the 
RG equations break down. This defines a new
intrinsic energy scale $\Delta$, below which scale invariance is 
lost and no further renormalization is possible.  
%This in turn also defines a 
%length scale $\xi = 1/\Delta$ beyond which the corresponding order parameter 
%$O$ no longer decays with a powerlaw, but remains constant 
%$O  \propto 1/\xi^{d}$. 
%For example, a small alternating field $h_{\rm alt}$ applied on a single chain 
%couples to the field $\mathbf{n}$ leading to a RG equation 
%$\dot h_{\rm alt}= (2-d_\mathbf{n})h_{\rm alt}$.  Breakdown occurs at an intrinsic 
%energy scale (gap) $\Delta \propto h_{\rm alt}^{1/(2-d_\mathbf{n})}$ and the order parameter 
%is  a staggered magnetization, with $\langle \mathbf{n} \rangle \propto 
%h_{\rm alt}^{d_\mathbf{n}/(2-d_\mathbf{n})}$ \cite{oshikawa1999, JPhysA.22.511}. 
%
\par 
The couplings  $\cpln$ and $\cple$  are the most relevant 
%(scaling dimension 1)  
and likely dominate. 
%A large  $\cpln$ corresponds to a N\'eel order along each chain and  a 
%large $\cple$ to a dimer order, while their sign selects between different N\'eel 
%or dimer patterns in the 2D lattice.  
A large negative $\cpln< 0$  corresponds to a regular N\'eel order 
while for a large positive value $\cpln>0$ the N\'eel order on each chain is collinear
to the neighboring chains.
A large $\cple < 0$ indicates a staggered dimer order on the
2D lattice while $\cple > 0$ gives a herringbone dimer structure.   
Depending on the microscopic couplings, it is also possible for the system to acquire 
a dimerized or magnetically ordered  ground state when one of the marginal  
couplings $\cpla$ or  $\cplb$, respectively,  reaches first the strong coupling limit 
\cite{PhysRevB.61.8871, PhysRevB.81.064432}.   Finally, 
the relevant coupling constant $\cplc$ with scaling dimension 
$d_{c}=3/2$   may also become largest, a  case  where no simple order can be 
assigned to the 2D system and a spin liquid phase is possible.  
\par 
We now proceed to numerically integrate the RG equations \eqref{rg-flow-a} 
for different lattice parameters until one of the couplings becomes unity
\cite{PhysRevB.81.064432,PhysRevB.54.9862}.
%This choice of cutoff for the integration is in fact somewhat arbitrary, but has
%given results which appear to compare well with 
%numerical data 
%Likewise, 
The constant $\const$ in the definition of the initial values in 
Eq.~(\ref{barecouplings}) must be of order unity according to bosonization, but cannot 
be derived exactly, so that we choose 
 $\const=1$ here.
We have also tested other values for the cutoff and $\const$ 
and found that the resulting phase diagram remains qualitatively the same, while
the quantitative positions of the phase transition lines may change slightly (see below).

\begin{figure}
\includegraphics[width=0.99\columnwidth,angle=0]{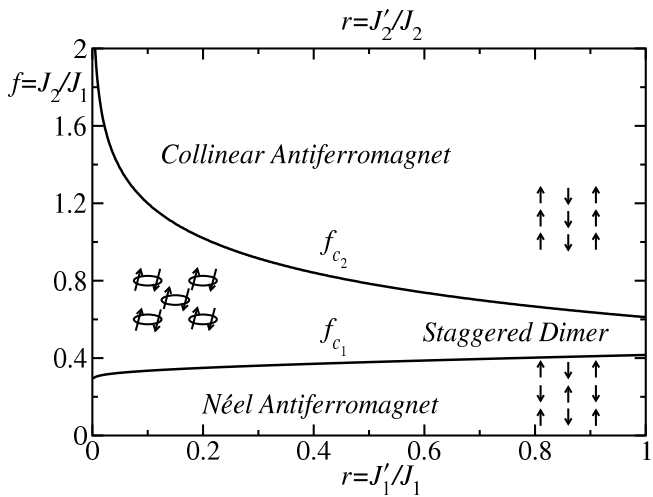}
\caption{ The phase diagram of the anisotropic $J_1$-$J_2$-$J_1'$-$J_2'$ 
model from the RG analysis as a function of the anisotropy $r$ and the frustration $f$.
The analytical functions in Eqs.~\eqref{fit-function-1} and \eqref{fit-function-2}
are indistinguishable on this scale.}\label{phase-diagram}
\end{figure}

There are two key parameters of physical interest in this model, 
namely the frustration $f=\frac{J_2}{J_1} =\frac{J_2'}{J_1'}$
and the anisotropy $r=\frac{J_1'}{J_1}=\frac{J_2'}{J_2}$, which determine the value of 
all four couplings \cite{footnote}.
The isotropic 2D model is recovered for $r\to 1$.
The resulting phase diagram as a function of $r$ and $f$ is
shown in Fig.~\ref{phase-diagram}.

{\it Lower critical line - }
The lower critical line $f_{c_1}(r)$
between the \emph{N\'eel} and the \emph{staggered dimer} order is determined by the 
competition between the relevant coupling
$\cpln<0$ and the marginally relevant coupling $\cpla>0$. 
The relevant coupling $\cple$ is initially zero and remains so small that 
it does not play a role here.
For a different choice of a strongly reduced value for $\const=0.5$ the critical line 
is slightly shifted down by 10\%, 
so that N\'eel order is slightly suppressed for $\const<1$ 
(and slightly stabilized for $\const>1$).

If we completely ignore all coupling constants except $\cpln$ and $\cpla$ 
in Eqs.~(\ref{rg-flow-a}), 
the lower critical line $f_{c_{1}}(r)$ can be analytically determined to be 
\BE\label{fit-function-1}
f_{c_{1}}(r) = J_c/J_1+\frac{a_1}{b_1 -\ln r} ~, 
\EE 
which is the critical value of frustration at which the breakdown energy scales
for dimer order $\Delta_a = \Lambda_0(1-1/\cpla)$ and N\'eel
order $\Delta_n=\Lambda_0 \cpln$ become equal.  In terms of the initial couplings
we have $a_1 \approx 1/1.723$ and $b_1 \approx 1 +\ln[2\pi v/(J_1-J_c)\const^2]$ 
from Eq.~(\ref{barecouplings}).
Remarkably, this analytical form remains a rather accurate estimate for the fully integrated
RG flow even in the presence
of all other couplings, albeit with fitted values of $a_1=0.459$ and  $b_1=2.629$.
In the limit $r \rightarrow 1$, we predict the critical point 
$f_{c_{1}}(1)\rightarrow 0.416$ for the transition to a staggered dimerized phase in 
the 2D $J_1$-$J_2$ square antiferromagnet. 
\par 

\par 
{\it Upper critical line - }
While a single chain remains in the dimer phase for $J_c<J_2<\infty$ \cite{PhysRevB.54.9862,
PhysRevB.55.299}, in the  $J_1$-$J_2$-$J_1'$-$J_2'$-model  a large $J_2$ can destroy 
the dimer phase \cite{PhysRevB.84.144407}. This behavior defines an upper critical line 
$f_{c_{2}}(r)$, with $f_{c_{2}}(0)\rightarrow \infty$, beyond which 
the system is driven out of the dimer phase. 
The RG Eqs.~(\ref{rg-flow-a})  {\it cannot}
 capture this behavior since it occurs for large deviations from $J_2=J_c$.
However, it is possible to consider the limit $J_2 \gg J_1$ as a new fixed point, 
where (unfrustrated) chains along the horizontal $J_2$ direction in Fig.~\ref{model} (bottom)
are weakly coupled by all other couplings. 
The low energy physics is determined by the competition of a coupling with $J_2'$
to every second chain (rectangular pattern) and a coupling $J_1$ or $J_1'$ which 
couples even and odd chains 
(zigzag pattern) \cite{PhysRevLett.98.077205,PhysRevB.84.174415,PhysRevB.54.9862,PhysRevB.55.299}. 
Here the weaker coupling 
$J_1'<J_1 $ can never dominate over $J_1$ in the course of the RG flow and can
safely be ignored. 
Analogously to the RG analysis above, the perturbation 
of the independent chain fixed point is given by 
%\BEA
%\delta \mathcal{H}'& =&  \delta \mathcal{H}_{\cplc=0}+ 
%2\pi v  \sum_{\nu}\Big[ 
%\lambda_{u} \left(\mathbf{J}_{\nu,L}\cdot\mathbf{J}_{\nu+1,R}+\mathrm{e.c.}\right) \nonumber \\
%&&\qquad  + \lambda_{w} \mathbf{n}_{\nu}\cdot\partial_x\mathbf{n}_{\nu+1}  +
%\lambda_{v}  \epsilon_{\nu}\partial_x\epsilon_{\nu+1}
%\Big]  ~. 
%\EEA
%
%
\BEA\label{hpertb}
\delta \mathcal{H}'& = & 2\pi v  \sum_{\nu}\Big[ 
\lambda_{a}  \mathbf{J}_{\nu,L}\cdot\mathbf{J}_{\nu,R}    + 
\lambda_{b}  (\mathbf{J}_{\nu,L}\cdot\mathbf{J}_{\nu+2,R}+\mathrm{e.c.})    \nonumber  \\
&+&\lambda_{\epsilon} \epsilon_{\nu} \epsilon_{\nu+2} 
+ \lambda_{n} \mathbf{n}_{\nu}\cdot\mathbf{n}_{\nu+2} 
+\lambda_{v}  \epsilon_{\nu}\partial_x\epsilon_{\nu+1} \\ 
&+& \lambda_{u} \left(\mathbf{J}_{\nu,L}\cdot\mathbf{J}_{\nu+1,R}+\mathrm{e.c.}\right)+  
 \lambda_{w} \mathbf{n}_{\nu}\cdot\partial_x\mathbf{n}_{\nu+1} \Big]  ~.  \nonumber 
\EEA
The initial bare couplings $\lambda_{\epsilon}$ and  $\lambda_{v}$ vanish, 
$\lambda_{a}\approx-0.36$ \cite{PhysRevB.54.R9612},  and  
\BE
\lambda_{u}=\frac{2J_{1}}{2\pi v}~,~~
\lambda_{w}=\const^{2}\frac{J_{1}}{2\pi v}~,~~
\lambda_{b}=\frac{J_2'}{2\pi v}~,~~
\lambda_{n}=  \const^2 \frac{J_2'}{2\pi v}~.  \nonumber 
\EE
The RG flow of this set of operators reads 
\begin{subequations}
\BEA \label{ma-strong-jnnn}
\dot \lambda_{a} &=& \lambda_{a}^{2} +\lambda_{\epsilon}^{2} - \lambda_{n}^{2}  
+\frac{1}{4} ( \lambda_{w}^{2}-\lambda_{v}^{2})  ~, \\
\dot{\cplb}  &=&   \cplb^2 -\cple\cpln
+ \cpln^2~, \\
\dot{\cple}  &=&  \cple +\frac{3}{2}\cpla\cple
-\frac{3}{2}\cplb\cpln  ~, \\
\dot{\cpln}  &=&  \cpln -\frac{1}{2}\cpla\cpln
-\frac{1}{2}\cplb\cple +\cplb\cpln  ~,  \\
\dot \lambda_{u} &=& \lambda_{u}^{2} 
-\frac{1}{2} \lambda_{v}\lambda_{w}+\frac{1}{2}\lambda_{w}^{2} ~,  \\
\dot \lambda_{v} &=&\frac{3}{4}\lambda_{a}\lambda_{v} 
-\frac{3}{2} \lambda_{u}\lambda_{w}~,  \\ 
\dot \lambda_{w} &=& -\frac{1}{4}\lambda_{a}\lambda_{w}  
-\frac{1}{2} \lambda_{u}\lambda_{v} 
+\lambda_{u}\lambda_{w}  ~, 
\EEA   
\end{subequations}
where the rectangular ($\cplb$, $\cple$, $\cpln$) and the triangular 
($\lambda_u$, $\lambda_v$, $\lambda_w$) patterns renormalize almost independent 
of each other, connected only indirectly via the in-chain marginal coupling $\lambda_{a}$. 
In this case, the equations result in a competition between 
a \emph{collinear} N\'eel order with 
dominant $\cpln>0$ (rectangular pattern) and a dimer order with a dominant $\lambda_{u}>0$
(zigzag pattern).  While the {collinear} N\'eel order is different from the 
regular N\'eel order discussed above, the dimer order is actually continuously 
connected to the staggered dimer order in the opposite
limit $J_2\agt J_c$ 
\cite{PhysRevB.54.9862, PhysRevLett.98.077205,  PhysRevB.77.205121, PhysRevB.61.8871},
as indicated in Fig.~\ref{phase-diagram}.

Again, ignoring all coupling constants except for $\lambda_{u}$ and $\cpln$, we
can find an analytic solution for the upper phase transition line $f_{c_2}(r)$
\BE\label{fit-function-2}
f_{c_2}(r)=a_2-b_2\ln r~, 
\EE
with $a_2 =(1+\ln 2\pi v/\const^2)/\pi v$ and $b_2 =1/\pi v $. Remarkably, the 
integrated RG flow from all couplings is again fitted well by the same form with
$a_2=0.606$ and $b_2=0.259$.
Therefore, the critical point for transition from dimer to collinear N\'eel order in the 
2D model is predicted to be  $f_{c_2}(1)=0.606$.  

%Considering the quantitative uncertainties in integrating the RG equations
%to coupling constants of order unity, 
We now turn to an independent
numerical check of the phase diagram of
a minimal model of
two zig-zag chains
with in-chain frustration $J_2$ and inter-chain coupling $J_1'$ and $J_2'$
using the
DMRG algorithm \cite{white92}.
This model can be used to study the stability of the different phases, but
quantitative agreement with the full 2D $J_1$-$J_2$-$J_1'$-$J_2'$-model
should not be expected.
% \cite{PhysRevB.77.205121, PhysRevB.72.104435}. 
To obtain the phase diagram we use the dimer order 
$O_D$ parameter 
\BE
 O_D  = \frac{1}{2}\Big|\big[h_1(l)-h_1(l+1)\big]+\big[h_2(l)-h_2(l+1)\big]\Big|, 
\EE
with $h_\nu(l)=\langle \mathbf{S}_{\nu,l}\cdot \mathbf{S}_{\nu,l+1}\rangle$ 
and the two rung entropy $\tilde S= \mathrm{Tr}\tilde\rho\ln \tilde\rho$ in terms of 
the reduced density matrix $\tilde\rho$ of two successive rungs 
\cite{PhysRevLett.96.116401,PhysRevB.77.205121}.
Possible anomalies in $\tilde S$ or its derivative with respect to some microscopic coupling 
 $\tilde S'$ indicate  a first or second order phase transition respectively
  \cite{Nature.416.608, PhysRevLett.93.250404, PhysRevB.78.224413, PhysRevB.77.214418}. 
Measurements are taken close to the center of a system with open boundary conditions and we
then apply finite size scaling as a function of both 
the number of DMRG states kept $M$ and the size of the system $L$ \cite{PhysRevB.86.075133}.
%For our simulations we have used system sizes up to $L=256$ rungs 
%with $M=400$ states per block 
%kept while we require the  ground state energy convergence to be  $10^{-8}$. 
%For each lattice size  
%the number of states $M$ is extrapolated using a scale invariant function as described 
%in Ref.~\cite{PhysRevB.86.075133}. Then a finite size scaling analysis is made
%using an exponential function close to the critical point and 
%polynomials in the dimer region \cite{PhysRevB.84.144407}. 
%
%
%
\begin{figure}
\includegraphics[width=0.99\columnwidth,angle=0]{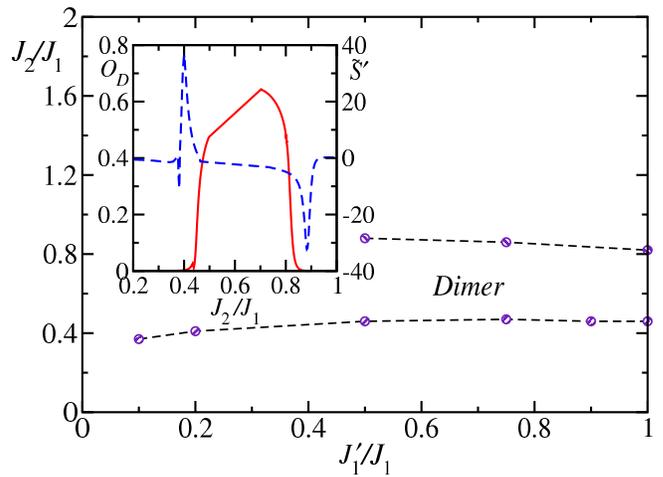}
\caption{Numerical DMRG results for the phase diagram of two chains with in-chain 
frustration $f=J_2/J_1$ and inter-chain couplings  $r=J_1'/J_1=J_2'/J_2$.
Inset: Dimer order parameter (red solid line) and 
the derivative of the two rung entropy (blue dashed line) as a function of 
$f$ for $J_1'=J_1$ and $L=128$ rungs.}\label{phase-diagram-dmrg}
\end{figure}
\par 
As shown in the inset of Fig.~\ref{phase-diagram-dmrg}, 
the onset of the dimer order $O_D$ coincides with the anomalies of the derivative 
$\tilde S'=d\tilde S/dJ_2$, indicating a second order phase transition.  
The resulting phase diagram as a function of %the analogous parameters
$r=J_1'/J_1$ and $f=J_2/J_1$ is shown in 
Fig.~\ref{phase-diagram-dmrg}. 
In the intermediate
region we find a non-zero order parameter $O_D$ corresponding to an in-chain 
dimerization and a {\it staggered} dimer arrangement in the 2D model.
Note that N\'eel order in the lower part of the phase diagram corresponds to 
ferromagnetically correlated rungs of the zig-zag chains and vice versa for the
collinear order.
The numerical data indicates that the intermediate phase is stable for all $r$.
%In the lower part of the phase diagram, antiferromagnetic correlations
%across the coupling $J_1'$ dominate, corresponding to a regular
%N\'eel order.  In the upper phase antiferromagnetic correlations dominate across
%the coupling $J_2'$, which corresponds to collinear N\'eel order.  

%
%
In summary, we have
considered the
general anisotropic $J_1$-$J_2$-$J_1'$-$J_2'$-model in Fig.~\ref{model}
which can be tuned from $r=0$ (decoupled chains) to $r=1$, where it becomes the
famous
$J_1$-$J_2$-model
\cite{anderson,balentsj1j2,becca1,gong,schulz,PhysRevB.42.8206,PhysRevLett.93.127202,PhysRevB.81.144410,PRB_79_024409,PRB_44_12050,PRB_85_094407,PRB_74_144422,sirker2006,PRL_91_197202,PhysRevB.78.214415,1309.6490,PRL_91_067201,PRB_86_045115,PRL_111_037202,PhysRevB.63.104420,PhysRevB.66.054401,PRB_41_9323,PRB_54_9007,PRL_78_2216,PRB_60_7278,PhysRevLett.91.017201,PRL_84_3173,PRL_87_097201,EPJB_73_117,PhysRevB.58.6394,1112_3331,PhysRevLett.97.157201,Senthil},
which is known to have an intermediate phase for $0.4 \alt J_2/J_1 \alt 0.6$ with
strongly debated underlying correlations.
%Early numerical works have shown
%For $r=1$ the range
%$0.4 \alt J_2/J_1 \alt 0.6$  of an intermediate phase has long ago been
%estimated in pioneering numerical works \cite{schulz}.
% and confirmed by all following studies,
%including modern coupled cluster
%methods \cite{PhysRevB.78.214415}.
%However, the nature of this phase (spin liquid or some kind of dimer or plaquette order)
%is still disputed.
One advantage of the $J_1$-$J_2$-$J_1'$-$J_2'$-model is that the
RG analysis becomes {\it exact} in the vicinity
of two fixed points in the phase diagram, namely close to $r\ll 1, \  f=J_c/J_1=0.241167$ and
close to $r\ll 1 \ll f$.  At those points a transition from a N\'eel order to a
staggered dimer order and a transition from a staggered dimer order to N\'eel order
can be rigorously shown. The phase transition lines behave asymptotically
according to Eqs.~(\ref{fit-function-1})
and (\ref{fit-function-2}), respectively.
For larger values of $r$ the
RG solution becomes less reliable, but the
predicted range of the staggered dimer phase
$0.416\lesssim J_2/J_1 \lesssim 0.606$ for $r=1$ is in surprising agreement with
previous studies.
In principle, the staggered dimer phase could also be limited for larger $r$, which
is in fact the case for the corresponding ladder model
without frustrating inter-chain couplings $J_2'=0$ \cite{PhysRevB.84.144407}.  However,
in the $J_1$-$J_2$-$J_1'$-$J_2'$-model there is a
different symmetry for large and small values of $f$, so an intermediate phase must
exist for all values of $r\leq 1$.  The most likely scenario is therefore that
the staggered dimer phase continuously extends from $r=0$ to $r=1$.
Indeed, our numerical simulations
and the RG equations do not indicate that the nature of this phases changes as a
function of $r$.  However, if the dimer phase does change it may do so
only partially as a function of $r$  and very recent numerical results indeed suggest
that for $r=1$
a spin liquid may exist {\it in part} of the phase diagram \cite{gong}.  Exploring the
full phase diagram of the $J_1$-$J_2$-$J_1'$-$J_2'$-model as a function of anisotropy
with 2D numerical methods is therefore a promising
future research direction.

%Another possibility is that at some value of $r$
%another transition to a competing intermediate phase takes place, but this scenario
%is unlikely from our numerical analysis on the corresponding ladder model, since
%we could find absolutely no change in behavior or instability as a function of $r$.
%There is also no indication from the RG equations that another competing coupling constant
%may become more relevant.  We therefore conclude that there is a staggered dimer order
%in the intermediate phase for all $r\leq 1$ and not a spin liquid.
%
\acknowledgments{We are thankful for valuable discussions with R. Dillenschneider,
J. Sirker, I. Affleck, and N. Sedlmayr.  This research was supported by
the Collaborative Research Center SFB/TR49 of the DFG.}

{%\newpage
\onecolumngrid
\appendix 
\section{Appendix: Field theory and RG flow}
Here, we sketch the derivation of RG equations [Eqs.~(6) and (9)].
%\eqref{rg-flow-a},  \eqref{la-strong-jnnn} and \eqref{ma-strong-jnnn}]. 
The formulation of the  relevant operators for the bosonization of the  
$J_1$-$J_2$-$J_1'$-$J_2'$-model can be done in terms of fermionic operators. 
The currents are expressed as 
\BE
J_\kappa^a (z_\kappa)  
= :\psi_{\kappa \eta}^{\dag}\frac{\sigma_{\eta \eta'}^a}{2}\psi_{\kappa \eta'}:(z_\kappa) ~,
\EE
where $\sigma^a$ are the Pauli matrices, $\eta$-s sum over the spin components 
($\eta=\uparrow,\downarrow$), and $\kappa$ denotes the chirality ($\kappa=R,L$ 
or $\kappa=+,-$ respectively). The chiral complex coordinates are $z_\kappa=-\kappa ix + v\tau$. 
The dimerization and staggered magnetization operators are given by \cite{PhysRevB.72.094416}
\BEA
\epsilon(z) &\sim& \frac{i}{2}\big[
:\psi_{R\eta}^{\dag} \psi_{L\eta}:(z)-:\psi_{L\eta}^{\dag} \psi_{R\eta}:(z) \big]~,\\
n^a (z) &\sim& \frac{1}{2}\sigma_{\eta\eta'}^a\big[ :\psi_{R\eta}^{\dag}\psi_{L\eta'}:(z)+ 
:\psi_{L\eta}^{\dag}\psi_{R\eta'}:(z)\big]~,   
\EEA 
where we simplify the notation using the variable $z$ for operators which  depend on both chiral variables $z_R$, $z_L$. 
\par  
The OPEs between $J_\kappa^a$, $\epsilon$,  and $n^b$ can be calculated using 
Wick's theorem \cite{cft} and the  two point correlation function 
\BE 
\langle \langle \psi_{\kappa \eta}(z_\kappa)\psi_{\kappa'\eta'}^\dag(w_{\kappa'})\rangle \rangle =
\delta_{\kappa\kappa'}\delta_{\eta\eta'}\frac{\gamma}{z_\kappa-w_\kappa}~, 
\EE
where $\gamma$ depends on the chosen normalization; here $\gamma=1/2\pi$.   
Keeping all relevant terms for the coupled field theories 
and after freezing out the charge degrees of freedom \cite{affleckleshouches}, 
the relevant OPEs are 
 \cite{PhysRevB.72.094416}  
\BEA\label{fundamentalopes}
J^a_\kappa(z_\kappa)J^b_{\kappa'}(w_{\kappa'})&=&\delta_{\kappa\kappa'}\left[ 
\frac{(\gamma^2/2)\delta_{ab}}{(z_\kappa-w_\kappa)^2} +i\epsilon_{abc}\gamma 
\frac{J_\kappa^c(z_\kappa)}{z_\kappa-w_\kappa}\right] \nonumber \\
J^a_\kappa(z_\kappa) \epsilon(w) & =&  i\kappa \frac{\gamma/2}{z_\kappa-w_\kappa} 
n^a(w) \nonumber \\ 
J^a_\kappa(z_\kappa) n^b(w) &=& i\frac{\gamma/2}{z_\kappa- w_\kappa} 
[\epsilon_{abc} n^c(w) -\kappa\delta_{ab} \epsilon(w)]\nonumber \\
\epsilon(z)\epsilon(w)&=& 
\frac{\gamma^2}{|z-w|}-|z-w|\mathbf{J}_R\cdot \mathbf{J}_L(w) \\ 
n^a(z)\epsilon(w)&=&-i\gamma|z-w|\left[ 
\frac{J_R^a(w_R)}{z_L- w_L}-\frac{J_L^a(w_L)}{z_R-w_R}
\right] \nonumber \\ 
n^a(z)n^b(w)&=& \frac{\gamma^2 \delta_{ab}}{|z-w|} 
+ |z-w|  \left[i\epsilon_{abc} \gamma \left[
\frac{J_R^c(w_R)}{z_L- w_L}+\frac{J_L^c(w_L)}{z_R-w_R}
\right] +\hat Q^{ab}(w)\right] \nonumber
\EEA 
with  $Q^{ab}$ denoting the zeroth order contraction between the fermionic fields. 
It tuns out that only the trace of this operator is relevant for the calculation of the 
coupled field theories' RG flow,  which after freezing out the charge degrees of freedom reads 
%\BE
%\hat Q^{ab} = \frac{1}{2} \sigma_{\eta\eta'}^a \sigma_{\tau\tau'}^b 
%\psi_{R\eta}^\dag \psi_{L\eta'} \psi_{L\tau}^\dag \psi_{R\tau'}
%\EE
\BE
\hat Q^{aa} = \mathbf{J}_R \cdot \mathbf{J}_L  ~. \nonumber 
\EE
\par 
The evolution of the bare couplings  is determined 
from the OPEs of the perturbing operators  [Eqs.~\eqref{hpert} and \eqref{hpertb}] 
using   \cite{cardy} 
\begin{equation}\label{couplingevolution}
\frac{d\lambda_k}{dl} =(2-d_k)\lambda_k-\frac{\pi}{v}\sum_{i,j} C_{ijk} \lambda_i \lambda_j~, \quad
\end{equation}   
where $d_k$ is the scaling dimension of the $O_k$ operator, $v$ the velocity, and $C_{ijk}$ 
the coefficient extracted from the OPE $O_i(z)O_j(w)\sim O_k(w)$.    
Comparing Eq.~\eqref{rg-flow-a} with Eq.~(43) in 
Ref.~\cite{PhysRevB.77.205121} where a related geometry was studied, one can notice
two discrepancies  in  the coefficients of the equations for the evolution of $\cplc$ and 
$\cpla$. For the former,  we believe the difference comes from the $1/2$ factor difference 
of the second term in the current-current OPE, cf.~Eq.~\eqref{fundamentalopes} and 
Eq.~(25) in Ref.~\cite{PhysRevB.77.205121}.  
For the latter, the difference arises from the symmetrization of the three chain geometry. 
A very illustrative example is the $\on \on$ OPE. From Eq.~\eqref{fundamentalopes} we have 
\BEA
\on(z)\on(w) & =&   \frac{\gamma^2}{|z-w|} \sum_\nu  \left( J^a_{R,\nu} J^a_{L,\nu}
+ J^a_{R,\nu+1} J^a_{L,\nu+1}\right)\nonumber \\ 
&-& \frac{2\gamma^2}{|z-w|} \sum_\nu \left( 
J^a_{R,\nu} J^a_{L,\nu+1}+ J^a_{R,\nu+1} J^a_{L,\nu}
\right)~.   \nonumber 
\EEA
For two coupled chains the summation should be omitted which would give a factor of two 
difference  of the  $\cpln^2$ contribution to $\dot \cpla$ and $\dot \cplb$.  However, 
the summation is required for the symmetrization of the coupled field theories
and therefore the two terms acquire the same coefficient.
}


\begin{thebibliography}{99}
\bibitem{footnote} In general the ratio $J_2'/J_1'$ could be taken as an 
additional independent parameter, but such an extended analysis is not of interest for
the 2D model.
\end{thebibliography}
\end{document}